\documentclass[aps,prl,floatfix,twocolumn,superscriptaddress]{revtex4-1}
\usepackage{amsmath,amssymb}
\usepackage{graphicx}
\usepackage{epsfig}
\usepackage{psfrag}
\usepackage[usenames]{color}
\usepackage{xcolor}
\usepackage{hyperref}
\usepackage{soul,color}
\usepackage{physics}
\usepackage{tabu}
\usepackage{multirow}
\usepackage[normalem]{ulem}
\hypersetup{colorlinks=true,citecolor={blue},linkcolor={blue},urlcolor={blue}}

\begin{document}

\title{Nonequilibrium theory of the photoinduced valley Hall effect}






\author{I.~Vakulchyk}
\affiliation{Center for Theoretical Physics of Complex Systems, Institute for Basic Science (IBS), Daejeon 34126, Korea}
\affiliation{Basic Science Program, Korea University of Science and Technology (UST), Daejeon 34113, Korea}

\author{V.~M.~Kovalev}
\affiliation{Rzhanov Institute of Semiconductor Physics, Siberian Branch of Russian Academy of Sciences, Novosibirsk 630090, Russia}


\author{I.~G.~Savenko}
\affiliation{Center for Theoretical Physics of Complex Systems, Institute for Basic Science (IBS), Daejeon 34126, Korea}
\affiliation{Basic Science Program, Korea University of Science and Technology (UST), Daejeon 34113, Korea}

\begin{abstract}
A recent scientific debate has arisen: Which processes underlie the actual ground of the valley Hall effect (VHE) in two-dimensional materials? 
The original VHE emerges in samples with ballistic transport of electrons due to the anomalous velocity terms resulting from the Berry phase effect. 
In disordered samples though, alternative mechanisms associated with electron scattering off impurities have been suggested: (i) asymmetric electron scattering, called skew scattering, and (ii) a shift of the electron wave packet in real space, called a side-jump.
It has been claimed that the side-jump not only contributes to the VHE but fully offsets the anomalous terms regardless of the drag force for fundamental reasons, and thus, the side-jump together with skew scattering become the dominant mechanisms.
However, this claim is based on equilibrium theories without any external valley-selective optical pumping, which makes the results fundamentally interesting but incomplete and impracticable.
We develop in this paper microscopic theory of the photoinduced VHE using the Keldysh nonequilibrium diagrammatic technique, and find out that the asymmetric skew scattering mechanism is dominant in the vicinity of the interband absorption edge.
This allows us to explain the operation of optical transistors based on the VHE.
\end{abstract}


\date{\today}

\maketitle



\section{Introduction} The concept of the Hall effect is the emergence of an electric current or other flux of particles in a sample in the direction transverse to both the dragging force and the external magnetic field, which should be finite for the effect to take place.
If similar phenomena happen in the absence of a magnetic field, they are referred to as \textit{anomalous Hall effects} (AHEs)~\cite{RefNagaosa}.
The eminent examples of the AHE include the Hall effect in magnetic materials (with their built-in sample magnetization), the spin Hall effect, where the role of the magnetic field is played by spin-orbit interaction, and the valley Hall effect (VHE)~\cite{XiaoVHE, Mak1489, RefOurVAE, Jin893}, which emerges in two-dimensional (2D) Dirac materials possessing nonequivalent valleys in reciprocal space, like transition metal dichalcogenide (TMDC) monolayers~\cite{RefTMDs1, RefTMDs2, RefLiu2019}.
There, electrons and holes occupy two valleys, K and K$'$, that are connected by time-reversal symmetry. 
TMDCs also represent a promising platform and testing ground for optoelectronics~\cite{RefXu2020, RefKang2020} and spin valleytronics~\cite{PhysRevLett.124.166803} as direct band gap materials that obey valley-dependent optical selection rules~\cite{PhysRevLett.99.236809, PhysRevB.77.235406}.
These properties make them fundamentally interesting and appealing for device design~\cite{RefLi2020}.

It is commonly accepted that there exist three principal mechanisms behind the AHE in non-magnetic materials~\cite{Dyakonov}: (i) the Berry phase stipulated anomalous velocity term (also called the intrinsic contribution)~\cite{RevModPhysXiao}, (ii) the side-jump contribution, and (iii) the skew scattering (asymmetric) contribution.
These three terms interplay and can partially compensate each other, as has been reported in recent works on electron~\cite{RefSinitsyn, GlazGolArEl} and exciton~\cite{GlazGolArEx} transport in semiconductors.
In particular, one important recent work~\cite{GlazGolArEl} shows that the side jump and skew scattering should not be disregarded under photon or phonon drag conditions, as is usually done when considering the VHE~\cite{XiaoVHE, RefOnga, RefOurVAE, RevModPhysXiao}.
More precisely, it has been demonstrated that the side jump compensates for the intrinsic contribution to conductivity, and moreover, some terms in the side jump survive.

These fundamental conclusions undoubtedly play an important role in our understanding of the microscopic processes underlying the VHE. 
However, the existing theories only consider equilibrium electrons initially occupying two nonequivalent valleys.  
The valley Hall currents resulting from these electrons flow in opposite directions and, being of the same magnitude, the currents cancel each other out, leaving zero-net VHE current in the sample.
To observe a nonzero valley Hall current in actual experiments~\cite{Mak1489}, the sample should be illuminated by an external circularly polarized electromagnetic field of light. 
This destroys the time-reversal symmetry and predominantly populates only one of the valleys due to the valley-dependent interband optical selection rules. As a result, the current contributions from nonequivalent valleys do not annul each other. 
It is important, then, to consider nonequilibrium photo-excited electrons since they are the ones actually contributing to the VHE. 
While this idea has been briefly mentioned in literature~\cite{PRB_Olsen}, it has not been rigorously studied. 
In the meantime though, the light-induced AHE has generally become an active field of research~\cite{RefMclver2020}.
Analysis~\cite{Mak1489} of experimental VHE observations is based on a phenomenological expression of the form $\sigma_\textrm{H}\propto\delta n$, where $\sigma_\textrm{H}$ is the valley Hall conductivity and $\delta n$ is the electron density imbalance between the valleys due to interband photogeneration. 
The standard derivation of this dependence is based on the Berry-phase-related expression applicable to ballistic samples.

In this paper, we analyze the applicability of this approach in the presence of all relevant electron-impurity scattering processes.
We pose an intriguing question: Do these statements (regarding the partial compensation of the intrinsic contribution) remain valid in the case of optically driven systems based on Dirac materials when circularly polarized light pumps one of the valleys?
The answer to this question is of utmost importance not only from a fundamental viewpoint (since 2D Dirac materials are prone to interact with light) but also from the perspective of optoelectronic applications, in particular, in novel van der Waals heterostructures.
We consider the intrinsic, side-jump, and skew scattering contributions to valley Hall photoconductivity using the nonequilibrium Keldysh diagram technique. 
Thus, we build a microscopic theory of the photoinduced VHE.
%
%
%
%
%
%

\section{General theory}
%
%
We consider a 2D system (Fig. 1) exposed to a circularly polarized light (which results in interband transitions),
\begin{eqnarray}
\textbf{A}(t)=\textbf{A} e^{-i\omega t}+\textbf{A}^*e^{i\omega t},
\end{eqnarray}
and thus $\textbf{A}=A(1,i\sigma)$ with $\sigma=\pm 1$,
and the in-plane alternating drag electric field is
\begin{eqnarray}
\mathcal{A}(t)
=
\mathcal{A}\left(e^{-i\Omega t}+e^{i\Omega t}\right),
\end{eqnarray}
where we assume that the drag field is linearly polarized so that $\mathcal{A}$ is real-valued.
At the end of the calculations, we will put $\Omega\rightarrow 0$ to find the static limit, which corresponds to the drag effect.
\begin{figure}[t!]
\centering
\includegraphics[width=0.48\textwidth]{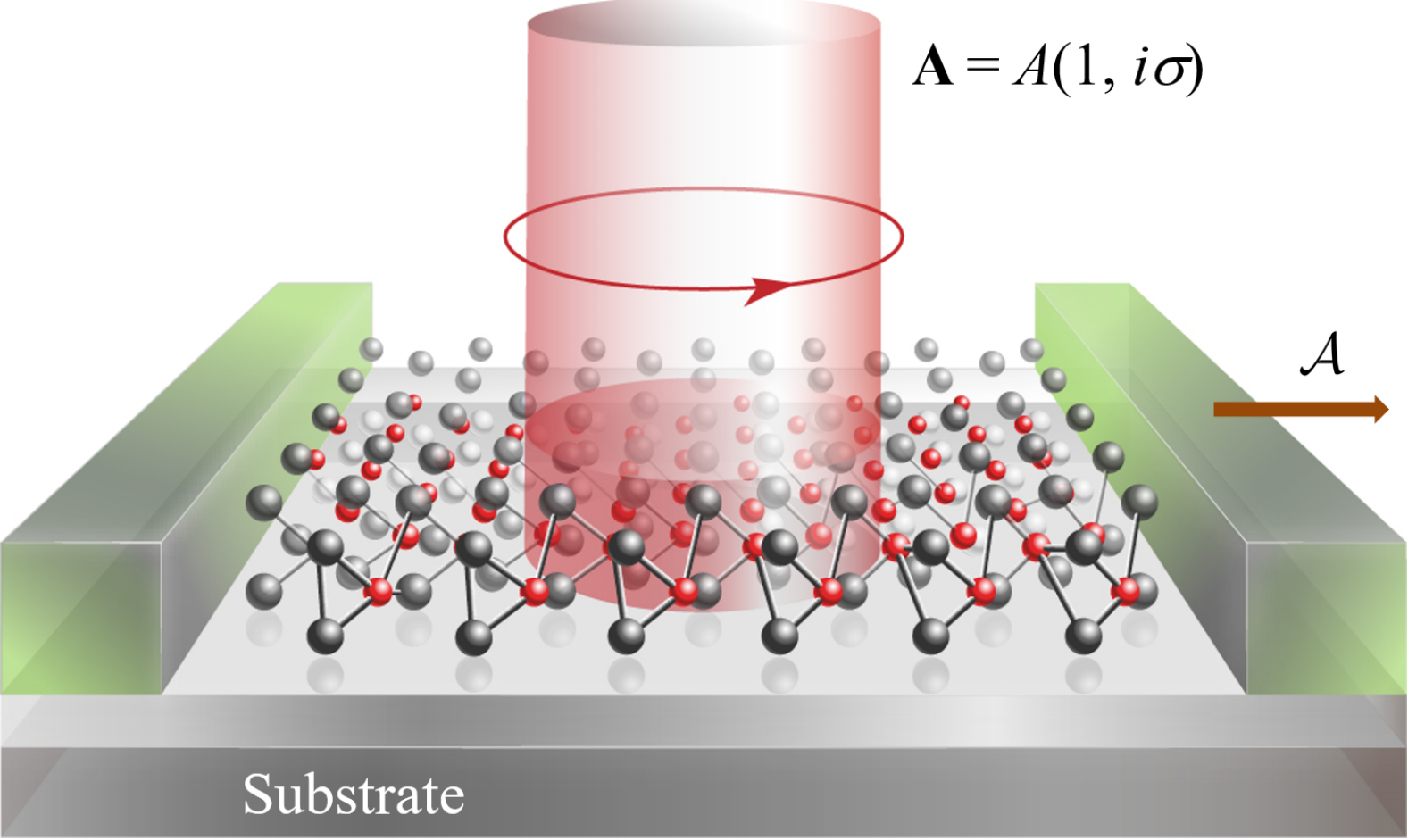}
\caption{System schematic of a 2D Dirac material exposed to circularly polarized light $\mathbf{A}$ and static drag field ${\cal A}$.
The light couples to the K or K$'$ valley depending on its polarization $\sigma$.}
\label{Fig1}
\end{figure}
We define the coordinates such that the drag  field is directed along the $y$ axis, and therefore our goal is to consider the valley Hall current along $x$. 
%
The full system Hamiltonian reads ($e<0$)
\begin{gather}
\label{EQ2}
H=\frac{\Delta}{2}\hat{s}_z+\textbf{V}\cdot\mathbf{p}-e\textbf{V}\cdot\mathbf{A}(t)-eV^y\mathcal{A}(t),
\end{gather}
where $\Delta$ is the monolayer material bandgap, $\mathbf{p}=p(\cos\phi,\sin\phi)$ is the electron momentum, $\textbf{V}=v_0(\eta \hat{s}_x,\hat{s}_y)$ is the velocity, $\eta=\pm1$ is the valley index, and $\hat{s}_\alpha$ are the Pauli matrices with $\alpha=x,y,z$. 
The Hamiltonian~\eqref{EQ2} is written in sub-lattice basis since the honeycomb lattice of a TMDC monolayer can be looked at as two triangle sub-lattices inserted into each other. 
However, it is instructive and physically transparent to work in the c- and v-band basis (the cv basis in what follows). 
In our case, the external fields in~\eqref{EQ2} are uniform in space, thus conserving the electron momentum (which, hence, can be considered as a complex number). To transform into the cv basis, we use a unitary operator that depends only on the electron momentum~\cite{RefOurNJP},
\begin{gather}
\label{EQ3}
U=\left(
    \begin{array}{cc}
      \cos(\theta/2) & \sin(\theta/2) \\
      \sin(\theta/2)e^{i\eta \phi} & -\cos(\theta/2)e^{i\eta \phi} \\
    \end{array}
  \right),
\end{gather}
where $\cos\theta=\Delta/2\epsilon_p$, $\sin\theta=\eta v_0 p/\epsilon_p$, and $\epsilon_p=\sqrt{\left(\Delta/2\right)^2+v_0^2p^2}\approx\Delta/2+p^2/2m$, where the electron effective mass is $m=\Delta/(2v_0^2)$ at small electron momenta, $v_0p\ll\Delta$. Using ${\cal H}=U^+HU$, we find
\begin{gather}
\label{EQ4}
{\cal H}={\cal H}_0
-e\textbf{v}\cdot\mathbf{A}(t)
-ev^y\mathcal{A}(t),
\end{gather}
where
\begin{gather}
\label{EQ5}
{\cal H}_0=\left(
                               \begin{array}{cc}
                                 \epsilon_\text{c}(p) & 0 \\
                                 0 & \epsilon_\text{v}(p) \\
                               \end{array}
                             \right),~~
\textbf{v}=\left(
                               \begin{array}{cc}
                                 \textbf{v}_\text{cc} & \textbf{v}_\text{cv} \\
                                 \textbf{v}_\text{vc} & \textbf{v}_\text{vv} \\
                               \end{array}
                             \right)
\end{gather}
are the bare Hamiltonian and the velocity operator in the cv-basis, with 
$\epsilon_\text{c}(p)\equiv\epsilon_p$ and $\epsilon_\text{v}(p)=-\epsilon_p$ (we will just write $\epsilon_\text{c}$ and $\epsilon_\text{v}$ in what follows, keeping in mind that they both depend on the absolute value of the momentum; we will also omit $\hbar$ in the expressions below but restore it in the final formulas). 

The valley Hall current, being the linear response to external drag field $\mathcal{A}$, reads~\cite{Mahan},
\begin{eqnarray}
\label{EQ6}
j_x(t)&=&\int\limits_{{\cal C}}dt'Q_{xy}(t,t')\mathcal{A}(t'),\\
\label{EQQxy}
&&Q_{xy}(t,t')=ie^2\textmd{Tr}\,\left[{v}^xG(t,t'){v}^yG(t',t)\right],~~
\end{eqnarray}
where ${\cal C}$ stands for the Keldysh contour, $\textmd{Tr}$ is the trace operator that should be taken over the bands, $Q(t,t')$ is the generalized conductivity representing a linear response function, and
\begin{gather}
\label{EQ7}
\Bigl[i\partial_t-{\cal H}_0+e\textbf{v}\cdot\mathbf{A}(t)\Bigr]G(t,t')=\delta(t-t')
\end{gather}
defines the matrix Green's function in the cv-basis. 
It should be stressed that this (matrix) Green's function accounts exactly for the external pumping field. 
We can also write Eq.~\eqref{EQ6} as
\begin{eqnarray}
j_x(t)=j_x^{(1)}(\Omega)e^{-i\Omega t}+j_x^{(1)}(-\Omega)e^{i\Omega t}.
\end{eqnarray}
The in-plane electric field is
$E_y(t)=-\partial_t\mathcal{A}(t)$, and the current can be found as $j_x(\Omega)=\mathcal{A}[Q_{xy}(\Omega,\omega)+Q_{xy}(-\Omega,\omega)]$. 
Also, we define $j_x(\omega)=\sigma_\mathrm{H}(\omega)E_y$. 
It is then possible to express the static (with respect to in-plane electric field $E_y$) valley Hall photoconductivity by the standard formula~\cite{Mahan}:
\begin{gather}
\label{EQ13}
\sigma_\mathrm{H}(\omega)=\lim_{\Omega\rightarrow0}\frac{Q_{xy}(\Omega,\omega)-Q_{xy}(-\Omega,\omega)}{2i\Omega}.
\end{gather}
%
%
%
As electron conductivity is due to particles from the v band being excited by the external field, to find Eq.~\eqref{EQ13}, we have to consider the Green's function of the electrons in the c band while accounting for interband pumping. This Green's function $G$ is the solution of Eq.~\eqref{EQ7} with the retarded (advanced) component $G^{R,A}$ and the lesser component $G^<(\varepsilon)=f_0(\varepsilon)\left[G^{A}(\varepsilon)-G^{R}(\varepsilon)\right]$, where $f_0(\varepsilon)$ is the stationary nonequilibrium distribution function of the c-band electrons under interband pumping. 
This stationary nonequlibirium electron distribution is characterized by the balance of electron generation and recombination. 
Thus, the retarded and advanced Green's functions read $G^{R,A}=(\varepsilon-\varepsilon_c\pm i/2\tau_i\pm i/2\tau_r)^{-1}$, where $\tau_i$ is the (intraband) electron momentum scattering time over impurities, and $\tau_r$ is the interband recombination time, or in other words, the lifetime of the electrons in the c-band. 

The nonequilibrium distribution function can be directly found from the equation of balance expressing the equality of generation and recombination processes in the form $f_0(\varepsilon)/\tau_r=g(\varepsilon)$, where the generation probability is $g(\varepsilon)=2\pi|M_\textrm{cv}(\bold{p}=0)|^2\delta(\varepsilon-\omega-\varepsilon_\textrm{v})$. 
Here, the interband matrix element, $|M_\textrm{cv}(0)|^2=|e\bold{v}_\textrm{cv}\bold{A}|^2=e^2v^2_0A^2(\eta+\sigma)^2$, is taken in the vicinity of the bottom of the c-band, $\bold{p}\approx0$. 
This regime is the most interesting for us since electrons find themselves in the c-band from optical absorption or scattering by impurities only, and thus we can disregard other sources of conducting electrons (such as the thermal ionization of shallow impurities). 
Then, the (vertical) optical transitions occur at very small electron momenta $p$. 
The factor $(\eta+\sigma)^2$ reflects the valley-selective interband optical rules for the circularly polarized pumping electromagnetic field.  
Finally, for the distribution function we find $f_0(\varepsilon)=2\pi\tau_r|M_\textrm{cv}(0)|^2\delta(\varepsilon-\omega-\varepsilon_\textrm{v})$. 
The same expression can be found by the Feynman diagrams technique~\cite{[{See Supplemental Material at [URL], which gives the details of the derivations of the main formulas}]SMBG}. 
Indeed, the bare self-energy of the photo-excited electrons in the c-band reads $\Sigma^<_\textrm{c}(\varepsilon)=|M_\textrm{cv}(0)|^2G_\textrm{v}(\varepsilon-\omega)$, as seen in Fig.~\ref{Fig2}(a). 
The ladder renormalization of this expression [Fig.~\ref{Fig2}(b)] gives the lesser Green's function $G^<_\textrm{c}(\varepsilon)=2\pi i f_0(\varepsilon)\delta(\varepsilon-\varepsilon_\textrm{c})$, where the distribution function $f_0(\varepsilon)$ has the same form as the one found from the equation of balance discussed above. 
Having defined the Green's functions describing the stationary nonequilibrium state, we can now analyze all the contributions to the photoinduced VHE. 
%
%
%
%
%
%
%
%
%
%
%
%
%
%
%

\section{Intrinsic contribution} 
\begin{figure}[t!]
\includegraphics[width=0.48\textwidth]{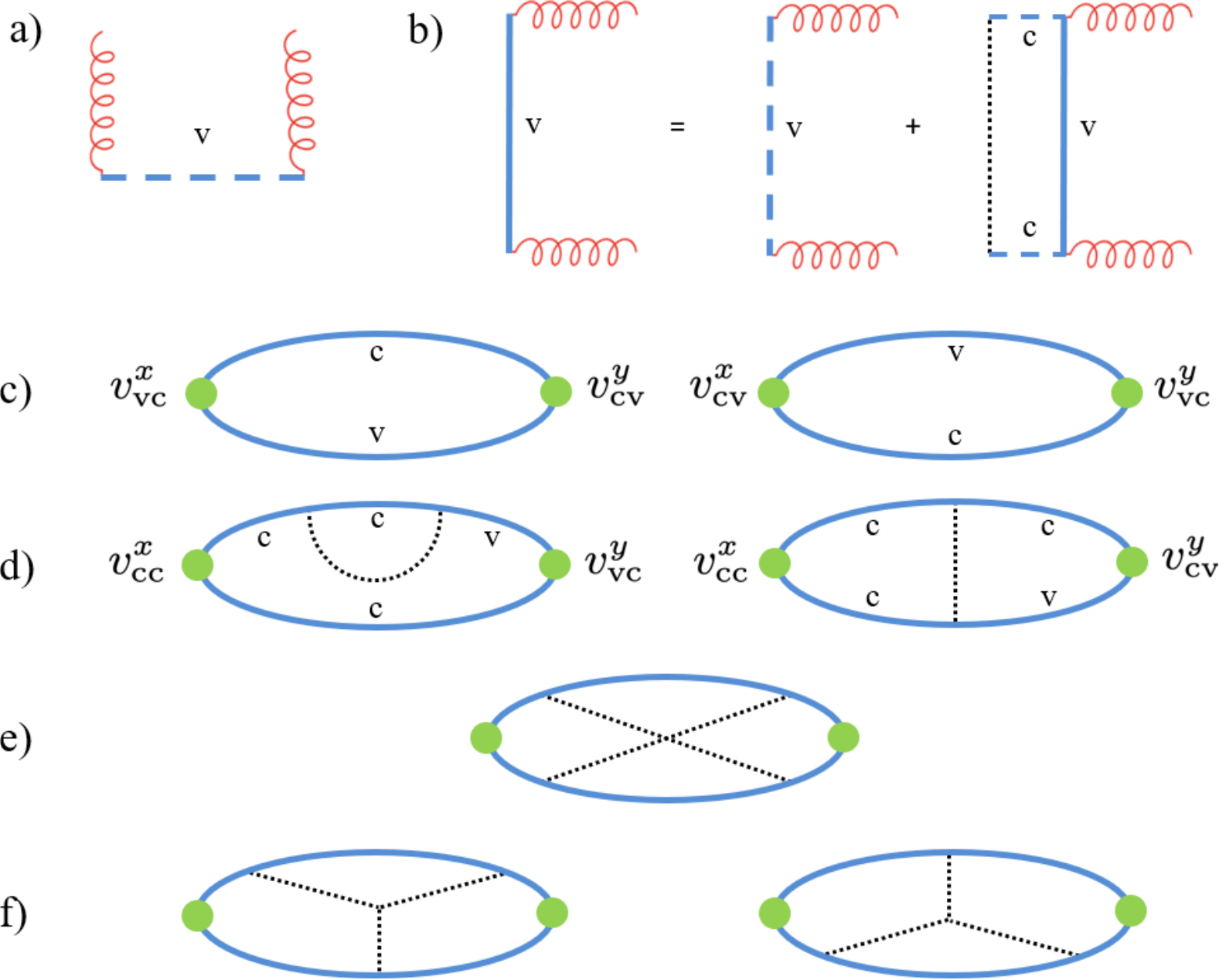}
\caption{
Feynman diagrams for the photoinduced VHE. (a) Bare self-energy of the photoinduced electrons in the conduction band. 
(b) Integral equation for the renormalized self-energy.
(c) The intrinsic contribution. (d) Examples of side-jump diagrams. 
(e) The X-diagram (coherent skew scattering) contribution.
(f) Anisotropic skew scattering diagrams. 
The red helixes stand for the external circularly polarized light $\mathbf{A}(t)$; $v^x$ and $v^y$ are the velocity vertices, c and v, respectively, mark the Green's functions of the electrons in the conduction and valence bands, and the dotted lines indicate impurity scattering.
}
\label{Fig2}
\end{figure}

The intrinsic contribution is associated with the Berry phase of the electrons in a given valley. 
It constitutes several diagrams of the kind depicted in Fig.~\ref{Fig2}(c). 
Each of these diagrams contains the interband matrix elements of velocity vertices $v^x$ and $v^y$. In our case, when the Fermi level is in the material bandgap, the contribution of these diagrams consists of two terms having a different physical meaning. 
The first one is also present in the equilibrium state and it is associated with the occupied v band possessing a topological nature. 
The second contribution is directly associated with the nonequilibrium state and is determined by the photo-excited electrons in the c band and holes in the v band.
Within the simple symmetric two-band model of Dirac bands in the MoS$_2$ monolayer, the holes' contribution has the same form and just doubles the result. 
Calculation of the terms illustrated in Fig.~\ref{Fig2}(c) gives~\cite{SMBG} (restoring $\hbar$)
\begin{gather}
\label{intrinsic}
\sigma^\textrm{(I)}_\textrm{H}=2\eta\frac{e^2}{\hbar} \left(\frac{\hbar v_0}{\Delta}\right)^2n_e,
\end{gather}
where $n_e=\sum_\bold{p}f_0[\varepsilon_\textrm{c}(\bold{p})]$ is the density of the photo-excited electrons in the c-band of a given valley. 
This contribution has the same structure as the one in the equilibrium case; the principal difference is that $n_e$ represents here the density of photo-excited electrons instead of the density of thermal-equilibrium electrons.

\section{Side-jump contribution} 
The diagrams representing the side-jump contribution contain the interband matrix element of electron-impurity scattering as depicted in Fig.~\ref{Fig2}(d). 
To calculate the conductivity due to the side-jump impurity process, let us first introduce the impurity potential in the cv basis. We assume an elastic scattering approximation and short-range impurities, and using the matrix element of the disorder matrix in the cv basis we find~\cite{RefSinitsyn} 
%
\begin{eqnarray}
\nonumber
&&u(\mathbf{p},\mathbf{p'})=
u^0(\mathbf{p},\mathbf{p'})
\left\{
[1-\sin^2\left(\frac{\theta}{2}\right)
(1-e^{i\eta(\phi'-\phi)})]\right.\\
\nonumber
&&~~~\left.
\times\frac{\hat s_0+\hat s_z}{2}
+
[1-\cos^2\left(\frac{\theta}{2}\right)
(1-e^{i\eta(\phi'-\phi)})]
\frac{\hat s_0-\hat s_z}{2}\right.\\
\label{EqImpurity2}
&&~~~~~~~\left.
+\frac{1}{2}\sin\theta(1-e^{i\eta(\phi'-\phi)})\hat s_x
\right\},
\end{eqnarray}
where $\hat s_0$ is the unity matrix, $\phi$ and $\phi'$ are the angles corresponding to the momenta $\mathbf{p}$ and $\mathbf{p}'$, respectively, and $\langle| u^0(\mathbf{p},\mathbf{p'})|^2\rangle=n_iu_0^2$ with $n_i$ being the density of impurities and $n_iu_0^2=(m\tau_i)^{-1}$. Here $\langle...\rangle$ stands for the averaging over the positions of impurities.
%
%
%
%
%
%
%

%
%
%
%
%
%
%
In the case of near-resonant pumping, renormalization of the vertices is negligible~\cite{SMBG}. 
Then, the impurity lines take the form of free Green's functions with an additional form-factor and give averages of the kind
\begin{eqnarray}
\nonumber
&&\overline{ u_\text{cc}(\bold{p},\bold{p}')
G_\alpha(\bold{p}',\delta t)
u_\text{cv}(\bold{p}',\bold{p})}\approx\\
\label{EqImpAv}
&&~\approx
\frac{1}{m\tau_i}
\int\frac{d\mathbf{p}'}{(2\pi\hbar)^2}
\left\{
\sin(\theta)\frac{1-e^{i\eta(\phi-\phi')}}{2}
\right\}
G_\alpha(\mathbf{p}',\delta t).~~~~~
\end{eqnarray}
%
Performing the diagram calculations taking into account the mass operator renormalization (see Fig.~\ref{Fig2}(b) and the Supplemental Material~\cite{SMBG}) we find 
\begin{eqnarray}
\label{sidejump}
\sigma_\textrm{H}^\textrm{(SJ)}=-4\eta\frac{e^2}{\hbar}\left(\frac{\hbar v_0}{\Delta}\right)^2\frac{n_e}{2\tau_i\gamma},
\end{eqnarray}
where $\gamma=(\tau_i+\tau_r)/2\tau_i\tau_r$ includes the recombination time $\tau_r$ reflecting the nonequilibrium nature of the effect. 
Again, the contribution of the side-jump process is determined by the density of the photo-excited electrons, $n_e$.

\section{Coherent skew scattering} \textcolor{black}{The skew mechanism is associated with asymmetric electron scattering by impurities. 
It should be described beyond the standard Born approximation in electron-impurity scattering probability. 
Coherent skew scattering involves pairs of closely located impurities, and can be illustrated by a diagram with crossed impurity lines [Fig.~\ref{Fig2}(e)]. 
In the framework of standard Drude theory, the diagrams possessing crossing impurity lines are parametrically small and usually do not play any role (except for the theory of weak localization, where maximally crossed diagrams are responsible for the effect). Nevertheless, the so-called $X$- and $\Psi$-type diagrams play an essential role in the AHE~\cite{AdoEPL2015, AdoPRB2017}. 
For a delta-correlated disorder, the contribution of the $\Psi$ diagram vanishes, leaving only the $X$ diagram for calculation (see Ref.~\cite{SMBG}):
}
\begin{eqnarray}
\label{CoherentSkew}
\sigma_\textrm{H}^\textrm{(X)}=2\eta\frac{e^2}{\hbar}\left(\frac{\hbar v_0}{\Delta}\right)^2\frac{n_e}{(2\tau_i\gamma)^2}.
\end{eqnarray}
\textcolor{black}{
Summing Eqs.~\eqref{intrinsic},~\eqref{sidejump}, and~\eqref{CoherentSkew}, we find }
\begin{eqnarray}
\nonumber
\sigma^\textrm{(I)}_\textrm{H}
+
\sigma^\textrm{(SJ)}_\textrm{H}
+
\sigma^\textrm{(X)}_\textrm{H}=2\eta\frac{e^2}{\hbar}\left(\frac{\hbar v_0}{\Delta}\right)^2n_e\left(1-\frac{1}{2
\tau_i\gamma}\right)^2
\\
\label{StandardSkewResult}
=2\eta\frac{e^2}{\hbar}\left(\frac{\hbar v_0}{\Delta}\right)^2n_e\left(1-\frac{\tau_r}{\tau_r+\tau_i}\right)^2.~~~~~~~~~~
\end{eqnarray}
\textcolor{black}{
In the limit $\tau_r\rightarrow\infty$ and assuming $n_e$ to be the equilibrium electron density, we recover the known result of the equilibrium VHE, when these three contributions cancel each other out~\cite{GlazGolArEl}. 
The typical values of the times are $\tau_r\sim \mu$s and $\tau_i\sim$ ps, i.e., $\tau_r\gg\tau_i$, allowing us to conclude that the mutual impact of these three contributions on the photoinduced VHE is negligibly small.
}

\section{Asymmetric skew scattering}
The last principal mechanism is associated with asymmetric electron skew scattering by impurities. 
It should also be described beyond the Born approximation~\cite{GlazGolArEl} and requires a non-vanishing impurity potential correlator of the third order. The corresponding Y-type Feynman diagrams, as in Fig.~\ref{Fig2}(f), have one-to-one correspondence with the Boltzmann equation result with an account of the asymmetric contribution to the electron-impurity collision integral~\cite{RefSinitsyn, GlazGolArEl, Sinitsyn2007}.
The calculation gives
\begin{eqnarray}
\label{ASkew}
\sigma_\textrm{H}^\textrm{(Y)}
=-\eta\frac{e^2}{\hbar}\left(\frac{u_0n_e}{\Delta}\right)\frac{\langle\varepsilon\rangle\tau_i}{(2\gamma\tau_i)^2\hbar},
\end{eqnarray}
where $\langle\varepsilon\rangle=n_e^{-1}\sum_{\bf{p}}(\varepsilon_p-\Delta/2)f_0[\varepsilon_c(\bold{p})]=(\hbar\omega-\Delta)/2$ is the mean energy of the photo-excited electrons in the c-band.
Evidently, the asymmetric skew scattering gives the dominant contribution to the photoinduced VHE, since the other contributions vanish, as we have shown above.


\section{Discussion} \textcolor{black}{Let us consider the approximations employed herein. First, photoinduced VHE conductivity contains the density of the photo-excited electrons,
}
%
\begin{eqnarray}
\label{limits}
n_e=\frac{m\tau_r|M_\textrm{cv}(0)|^2}{2\hbar^3}\Theta[\omega-\Delta],
\end{eqnarray}
where $\Theta[\omega]$ is the Heaviside step-function.
This formula has been derived by treating the external circularly polarized pump field as a perturbation. 
This is only valid as long as $\tau_r|M_\textrm{cv}(0)|/\hbar\ll1$.
In the opposite regime, $\tau_r|M_\textrm{cv}(0)|/\hbar\gg 1$, the pumping field cannot be considered as a perturbation. 
While a full theory of the photoinduced VHE in the strong-coupling regime is still missing, the intrinsic contribution (which dominates in ballistic samples) has been studied~\cite{RefOurNJP}.

The second limitation concerns the electron-impurity scattering: $\hbar/\tau_i$ should be small in comparison with the mean value of electron energy. 
In the case of the stationary nonequilibirum VHE, the characteristic energy of the photo-excited electrons is $\langle\varepsilon\rangle$, and thus $(\hbar\omega-\Delta)\tau_i/\hbar\gg1$ should be fulfilled.

Furthermore, we have accounted for one principal mechanism resulting in the establishment of a stationary nonequilibrium state of photo-excited electrons: the interband recombination. 
In principle, there also exist other mechanisms limiting electron lifetime in the band. 
Among them is intervalley electron scattering via impurities or phonons, where such transitions require large values of electron momentum transfer resulting in long electron lifetimes (comparable with $\tau_r$). 
Another mechanism is electron capture by impurities. These phenomena may play an important role in the photoinduced VHE and require a separate consideration.


Our results suggest an explanation of the operation of an optical transistor based on the VHE reported in an experimental work~\cite{Mak1489}.
There, the authors show a quasi-linear dependence of photoconductivity $\sigma_\textrm{H}$ on the density of the photoinduced electrons $n_e$.
It is demonstrated that the slope of the curve $\sigma_\textrm{H}(n_e)$ is controlled by the gate voltage, and the change of the inclination angle cannot be explained by the equilibrium formula for the intrinsic photoconductivity (see Fig.~3 in Ref.~\cite{Mak1489} and the corresponding discussion). 
Neither such (experimental) behavior of $\sigma_\textrm{H}(n_e)$ can be interpreted by the side-jump contribution, in either equilibrium or nonequilibrium cases.
However, our formulas Eqs.~\eqref{ASkew} and \eqref{limits} provide a possible interpretation of the experimental behavior as we show that the main contribution to the photoconductivity originates from skew scattering. 
Indeed, the slope of $\sigma_\textrm{H}(n_e)$ is proportional to the mean energy of the photo-excited electrons $\langle\varepsilon\rangle=(\hbar\omega-\Delta)/2$. 
Since with the increase of the gate voltage the c-band becomes more populated, the chemical potential $\mu$ (or, more precisely, the electron quasi-Fermi level, which depends on the gate voltage) enters the band. 
Then, the Moss-Burstein effect~\cite{BM1, BM2} results in a shift $\Delta\rightarrow2\mu$ (where $\mu$ is measured from the middle of the gap) as some low-energy states in the c-band become occupied. 
Hence, the ratio $\sigma_H/n_e\propto\langle\varepsilon\rangle$ becomes gate-voltage dependent.
Moreover, the increase of the electron density results in an enhancement of the scattering processes.
These arguments elucidate the role that skew scattering plays in the VHE in nonequilibrium situations and support our theoretical findings. 
%


\section{Conclusions} 
We have developed a microscopic theory of the photoinduced VHE in two-dimensional Dirac materials by employing the Keldysh nonequilibrium diagrammatic technique and analyzing the impurity scattering mechanisms under nonequilibrium conditions.
We have demonstrated that while in ballistic samples the intrinsic Berry phase-related term is dominant, in disordered samples the main contribution to the Hall photoconductivity stems from asymmetric skew scattering, with the other principal contributions canceling each other out.
In this way, one is able to explain the operation of optical transistors based on the VHE.

We thank M.~Glazov, L.~Golub, and A.~Parafilo for useful discussions and important advice, J.~Rasmussen (RECON) for a critical reading of our paper, and E.~Savenko for help with the figures.  
We have been supported by the Institute for Basic Science in Korea (Project No.~IBS-R024-D1) and the Russian Science Foundation (Project No.~17-12-01039).

\bibliography{library}
\bibliographystyle{apsrev4-1}
\end{document}